\documentclass{aastex}
\usepackage{emulateapj5, psfig, epsfig}
\usepackage{apjfonts}

\begin{document}

\slugcomment{The Astronomical Journal, in press}

\received{}
\revised{}
\accepted{}

\lefthead{M. Catelan et al.}
\righthead{}

\singlespace

\title{Age as the Second Parameter in NGC~288/NGC~362?\\
II. The Horizontal Branch Revisited}

\author{M.~Catelan\altaffilmark{1,}\altaffilmark{2}}

\author{M.~Bellazzini\altaffilmark{3}}

\author{W.~B.~Landsman\altaffilmark{4}}

\author{F.~R.~Ferraro\altaffilmark{3}} 

\author{F.~Fusi Pecci\altaffilmark{3,}\altaffilmark{5}}

\and

\author{S.~Galleti\altaffilmark{3}}

\altaffiltext{1}
      {
       University of Virginia,
       Department of Astronomy,
       P.O.~Box~3818,
       Charlottesville, VA 22901-2436, USA;
       e-mail: catelan@virginia.edu
       }
\altaffiltext{2}{Hubble Fellow.}
\altaffiltext{3}
      {
       Osservatorio Astronomico di Bologna,
       via Ranzani~1,
       40127, Bologna, Italy;
       e-mail: bellazzini, flavio, ferraro, l\_galleti@bo.astro.it
      }
\altaffiltext{4}
      {
       Raytheon ITSS, NASA Goddard Space Flight Center, 
       Laboratory for Astronomy and Solar Physics, 
       Greenbelt, MD 20771, USA;
       e-mail: landsman@mpb.gsfc.nasa.gov
       }
\altaffiltext{5}
      {
       Stazione Astronomica di Cagliari, 
       Loc. Poggio dei Pini, Strada 54, 09012 Capoterra (CA), 
       Italy
      }
\begin{abstract}
We revisit the ``second parameter'' pair of globular clusters NGC~288/NGC~362 
on the basis of theoretical models for red giant branch (RGB) and horizontal 
branch (HB) stars. The results of the most extensive set of RGB/HB simulations 
computed so far for these clusters are presented for two different 
metallicities. Using several different analytical mass loss 
formulae for RGB stars, we derive relative ``HB morphology ages.''  
We compare them with the relative 
main-sequence turnoff ages derived by application of the ``bridge test'' by 
Bellazzini et al. (2001),  who
found that NGC 288 is $2 \pm 1$ Gyr older than NGC 362.  
We find that adoption of a higher metallicity (${\rm [Fe/H]} \approx -1.2$), 
as favored by the Carretta \& 
Gratton metallicity scale, makes age a much more plausible second parameter 
candidate for this pair than is the case when a lower metallicity 
(${\rm [Fe/H]} \approx -1.5$), closer to the Zinn \& West scale, is adopted.
However, while the different HB morphology of these two clusters can be 
reproduced by canonical HB models with ${\rm [Fe/H]} \approx -1.2$ and  
an age difference of 2~Gyr, this explanation is not without difficulty. 
In particular, we confirm previous suggestions 
that canonical models are unable to reproduce the detailed HB morphology of 
NGC~288 at its red end, for as yet unknown reasons. Finally, we show that the 
mass dispersion on the HB of NGC~362 is substantially larger than for NGC~288,  
and suggest that there is a correlation between the mass dispersion on the HB 
phase and the central density of globular clusters. This is presumably related 
to the way environmental effects affect RGB mass loss---another major second 
parameter candidate. We argue that, if confirmed, this central density--HB 
mass dispersion correlation will have to be taken into account in order to 
conclusively determine whether age may be considered the (sole) second 
parameter of HB morphology for this (and other) second parameter pair(s). 
\end{abstract}


\section{Introduction}
NGC~288 (C0050-268) and NGC~362 (C0100-711) form what is perhaps the 
best-known ``second parameter pair'' of globular clusters---NGC~362 
presenting a very red horizontal branch (HB), and just the opposite
occurring in the case of NGC~288. Given that the two globulars have 
very similar chemical composition (Shetrone \& Keane 2000), the 
question naturally 
arises: What is (are) the reason(s) for the dramatically different 
HBs of these two clusters? 

Demarque et al. (1989) estimated an age difference amounting 
to $\sim 5-6$~Gyr from analysis of the main-sequence 
turnoffs (TO) of the two clusters. Interestingly, just such an age 
difference appeared to be required, according to the authors' 
theoretical HB simulations, to account for the relative HB types of 
NGC~288 vs. NGC~362. Hence these authors argued that age is the 
second parameter for this pair. 

However, since that time many other analyses have 
been published favoring much smaller TO age differences between
these clusters (see Bellazzini et al. 2001, hereinafter Paper~I, 
for up-to-date references); 
in a few cases, the possibility that they might differ in age by less 
than 1~Gyr (Grundahl 1999; VandenBerg 2000), or even be coeval 
(VandenBerg \& Durrell 1990; Stetson, VandenBerg, \& Bolte 1996), 
was raised. Therefore, in view of the original claims by Demarque et 
al. (1989) that an age difference as large as 5~Gyr is {\em needed} to 
account for the HB morphologies of this second parameter pair, such 
recent TO age difference estimates would essentially rule out age as 
the (sole) second parameter---at least in the case of NGC~288/NGC~362. 

On the other hand, several theoretical analyses have also been carried 
out since Demarque et al. (1989) which relaxed significantly the constraints 
on the age difference that is required to account for this second parameter 
pair. While Lee, Demarque, \& Zinn (1988) reported an HB morphology-based 
age difference between NGC~288 and NGC~362 of 5.7 to 7.3~Gyr (cf. their 
Table~III), Lee, Demarque, \& Zinn (1994) revised these 
values to the range 3--4~Gyr. The main reason for this difference 
lies in the adopted absolute age: While Lee et al. (1988) adopted 
an age 18--19~Gyr for NGC~288, Lee et al. (1994) reduced this 
value to 14.9~Gyr. In fact, Catelan \& de Freitas Pacheco 
(1993, 1994) pointed out that age differences lower than 3~Gyr might 
be consistent with age as the sole second parameter for this pair, 
provided both clusters are younger than $\sim 10$~Gyr. In addition, 
as pointed out by Lee et al. (1994) and Catelan \& de Freitas 
Pacheco (1995), age differences based on HB morphology may decrease 
further when formulations for mass loss on the red giant branch  
(RGB) are employed which imply an increase in overall mass loss with 
age. Just such a behavior is indeed found in 
the case of the well-known analytical mass loss formula proposed 
by Reimers (1975a, 1975b), as first pointed out by Fusi Pecci \& 
Renzini (1975, 1976).  

More recently, Catelan (2000) has readdressed the subject of analytical 
mass loss formulae for red giant stars and their impact upon HB morphology. 
Not only did he point out that Reimers' (1975a, 1975b) formula still lacks  
a sufficiently strong empirical basis, but also that 
there are several other analytical mass loss formulae in the literature 
that can be equally well justified in terms of the currently available 
data. Importantly, Catelan found that each such formula impacts HB  
morphology in a different way. Hence previous HB morphology-based 
age difference estimates for second parameter pairs such as 
NGC~288/NGC~362 may require revision due to the uncertainty in the 
treatment of mass loss in RGB stars. 

Motivated by this, and in view of the new results from Paper~I
for the TO age difference between NGC~288 and NGC~362 based on the  
so-called ``bridge test'' ($\Delta t = 2 \pm 1$~Gyr), we have decided 
to investigate anew whether age might be considered the (sole) second 
parameter for this pair. 

In the next section, we present the theoretical framework adopted in 
this analysis. Specifically, 
in \S2.1 we tackle the HB morphology of NGC~362, while \S2.2 is devoted 
to the NGC~288 case. In \S3, the HB morphology-based age differences 
implied by application of each of Catelan's (2000) mass loss formulae 
are provided for two different assumptions on the metallicity of 
the pair. Comparison with the TO age difference estimate in Paper~I 
is provided in \S4, where the possibility that age may be the 
(sole) second parameter for this pair is critically discussed. 
We close the paper in \S5 by addressing 
the possible existence of a correlation between the central densities  
of globular clusters and the mass dispersion on the HB phase.

\vskip 0.15in
\section{Theoretical Framework}
The basic theoretical framework adopted in this paper (HB and RGB
evolutionary tracks, color transformations, synthetic CMD generation) 
is very similar to that used in Catelan (2000) and  Catelan, Ferraro, 
\& Rood (2001); we refer the interested reader to those papers and to 
the references quoted therein for details. In the present study, 
stellar evolutionary tracks from Catelan et al. (1998) and from 
Sweigart \& Catelan (1998) for $Z = 0.001$ and $Z = 0.002$, 
respectively, have been employed. Unless otherwise stated, a main 
sequence helium abundance $Y_{\rm MS} = 0.23$ was adopted in all cases. 

We assume, as a working hypothesis, that age is the second parameter 
of HB morphology. We also assume, in line with Paper I (see the extensive
discussion in \S2.1 therein, but also \S4 below), 
that NGC~288 and NGC~362 constitute a ``bona fide'' second-parameter pair, 
thus adopting the same metallicity for the two clusters. 
The underlying zero-age HB (ZAHB) mass distribution is approximated by 
a Gaussian deviate (see Catelan et al. 1998 for a detailed discussion). 
The instability strip edges are the same as adopted in Catelan et al. 
(2001). ``Observational scatter'' was included by means of a suitable 
analytical representation of the photometric errors from Paper~I. 

Because we are analyzing the ``horizontal'' distribution of HB stars,  our
derived mean HB mass and dispersion are little affected by the adopted distance
moduli, though there is a small dependence on the adopted reddenings.    We use
the constraints from Paper I, where we    derived a relative distance modulus
between NGG~288 and NGC~362 of $\Delta (M-m)_V =  +0.005 \pm 0.087$
(where NGC~288 has the slightly larger distance modulus). Consistent with these
constraints, we adopt for our simulations an identical distance modulus for
both NGC~288 and NGC~362, with the actual value [$(m-M)_V = 14.79$ for 
$Z = 0.001$ and $(m-M)_V = 14.70$ for $Z=0.002$] chosen to provide a best-fit 
to the observed ZAHB level.  For NGC~362, we use the reddening value  of 
$E(V-I) \simeq 0.042$~mag as implied by the Schlegel, Finkbeiner, \& Davis 
(1998) maps. For NGC~288 we adopt  $E(V-I) \simeq 0.039$~mag 
[$E(\bv) \simeq 0.03$~mag], in excellent agreement with Harris (1996)  and 
Ferraro et al. (1999). Within the errors, these reddening values are also 
in reasonable agreement with the relative reddenings inferred from Paper~I, 
namely: $\Delta E(V-I) = +0.015 \pm 0.014$. 
Note that the main results of the present paper, summarized 
in Fig.~7 below,  were found to be largely insensitive to the adopted 
reddening values.

%
\begin{figure*}[ht]
 \figurenum{1}
 \centerline{\epsfig{file=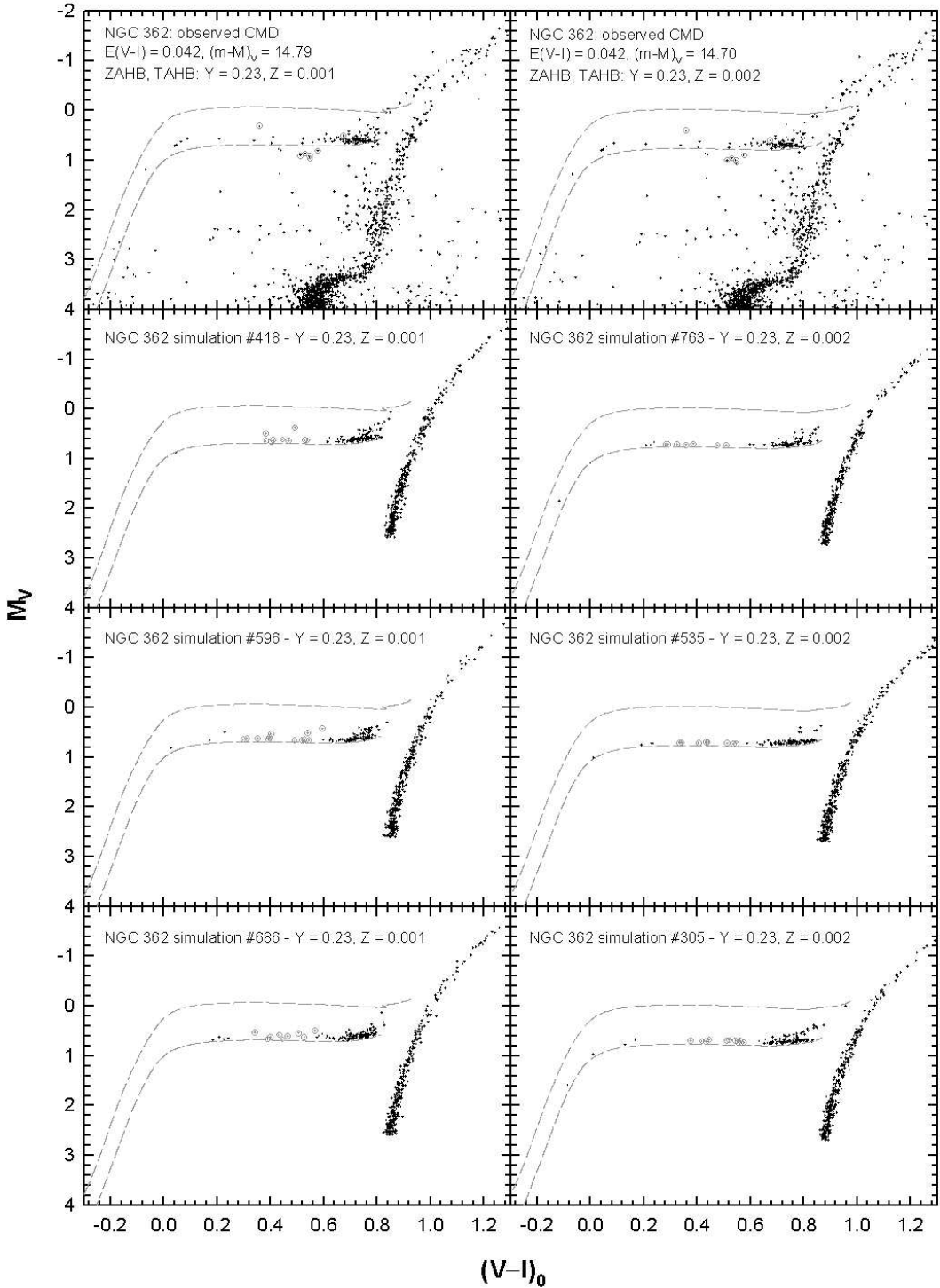,height=8in,width=5.75in}}
 \caption{The observed CMD of NGC~362 ({\em top row}, from Paper~I) 
          is compared against theoretical simulations for two 
          different metallicities: $Z = 0.001$ ({\em left column}) 
          and $Z = 0.002$ ({\em right column}). The observed CMDs 
          were dereddened by the indicated amounts, and shifted  
          by the distance moduli that brought best agreement 
          between the theoretical ZAHB and the lower envelope of the 
          red HB. 
          RR Lyrae variables are indicated by encircled dots;  
          their magnitudes are based on a single observation (and 
          thus do not represent an average). 
          Overplotted on all panels are the theoretical 
          ZAHB and TAHB lines.  
         }
\end{figure*}

%
\begin{figure*}[ht]
 \figurenum{2}
 \centerline{\epsfig{file=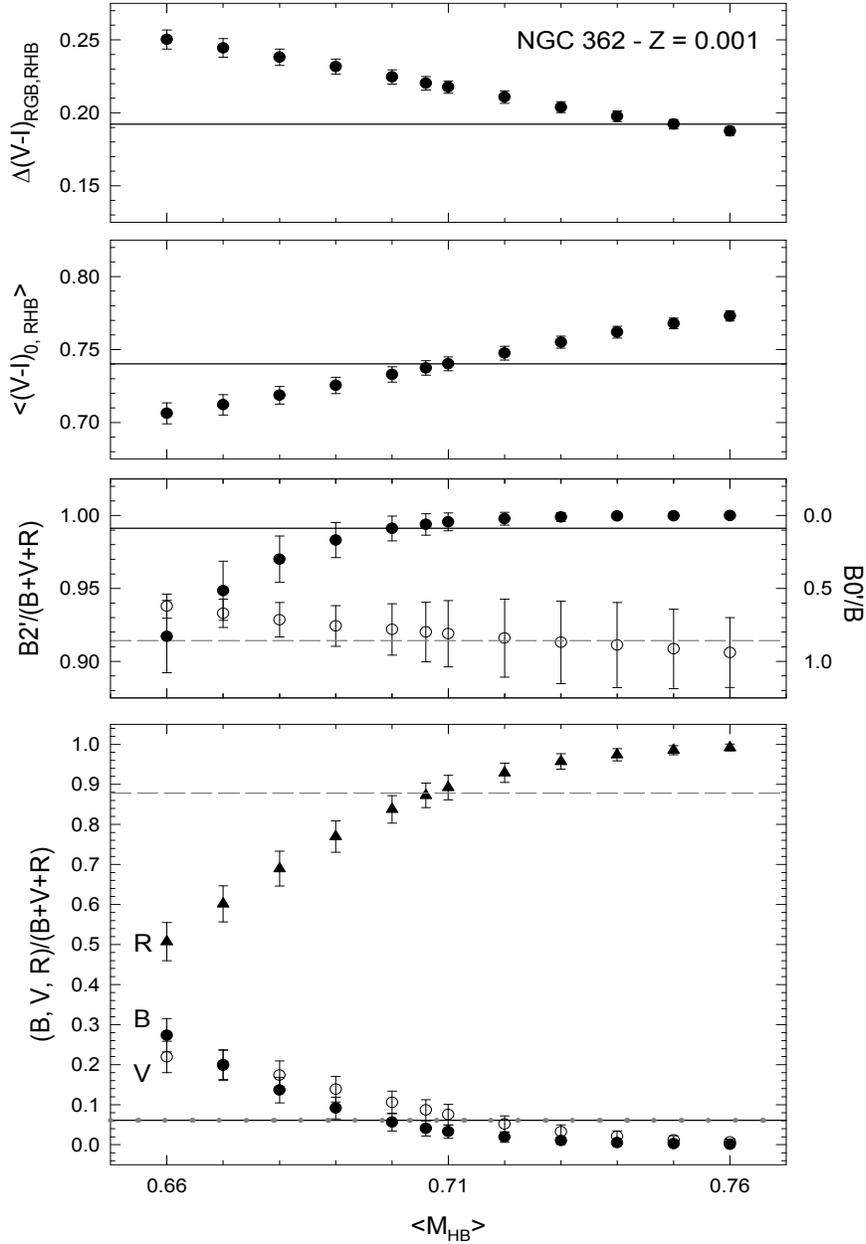,height=6.5in,width=4.5in}}
 \caption{Illustration of the dependence of several HB morphology 
          parameters on $\langle M_{\rm HB}\rangle$ for NGC~362 
          ($Z = 0.001$ case). In this example, the mass dispersion 
          parameter was held fixed at a value 
          $\sigma_M = 0.040 \, M_{\odot}$. The upper panel gives 
          the difference in color between the red HB and the RGB 
          at the level of the red HB, and the second panel from  
          the top shows the mean color of the red HB. In the third 
          panel from the top, the open circles and the gray dashed 
          line refer to $B0\arcmin/B$ (note the inverted scale on 
          the right), whereas the filled circles and solid line 
          refer to $B2\arcmin/(B+V+R)$.  
         }
\end{figure*}

%
\begin{figure*}[ht]
 \figurenum{3}
 \centerline{\epsfig{file=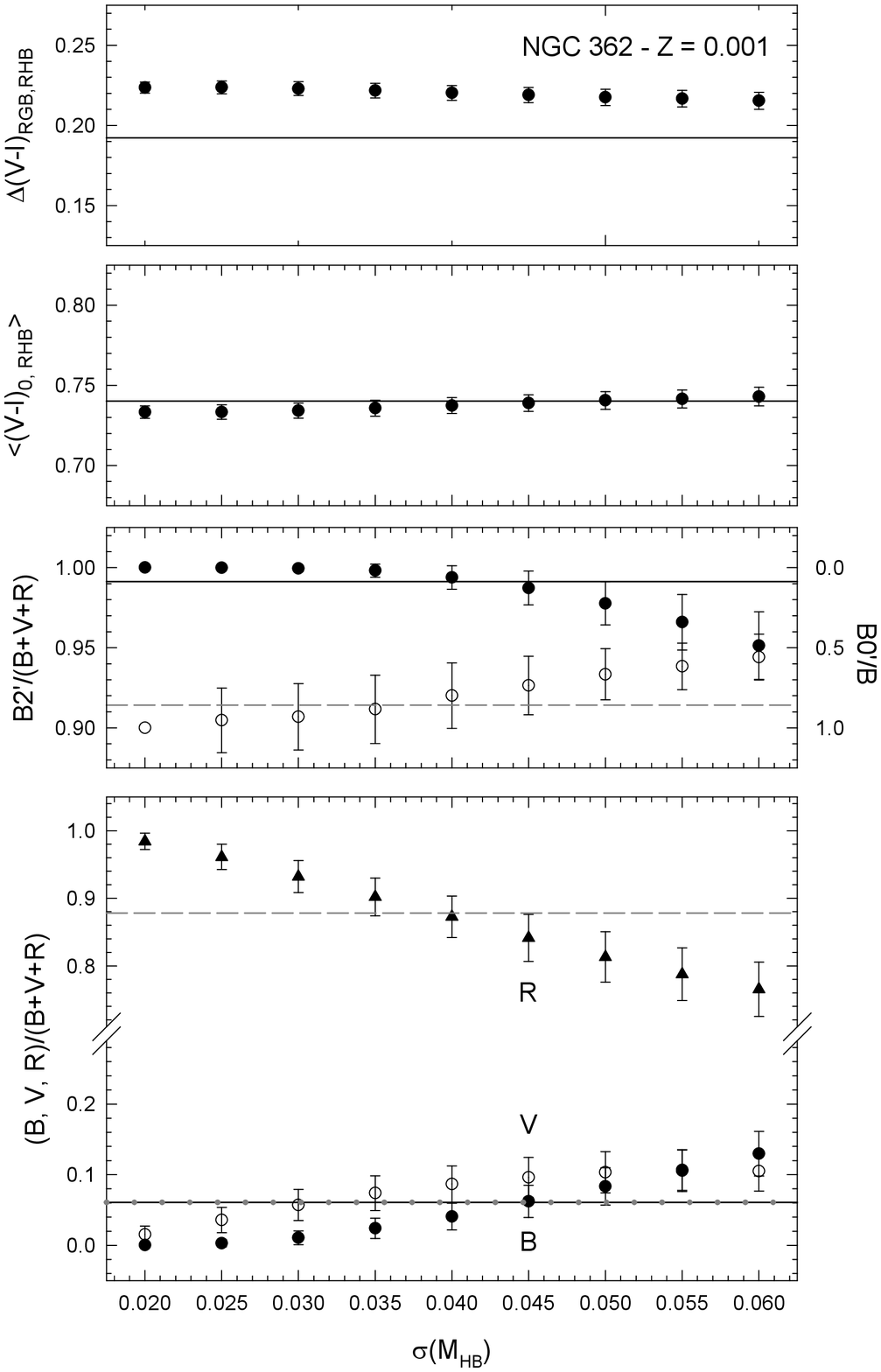,height=6.5in,width=4.5in}}
 \caption{As in Fig.~2, but for variations in $\sigma_M$. 
         }
\end{figure*}

\vskip 0.15in
\subsection{The Case of NGC~362}

\subsubsection{New HB Morphology Parameters: Definition}
Most previous studies of the pair NGC~288/NGC~362 have primarily 
utilized HB morphology parameters involving the overall 
numbers of blue, red and variable (RR Lyrae) HB stars ($B$, $R$, $V$, 
respectively). While such parameters can be very 
useful in some cases, one drawback associated with this approach is that 
parameters such as $(B-R)/(B+V+R)$ (the ``Lee--Zinn parameter'') tend to 
completely lose sensitivity to HB morphology variations for very blue (or 
very red) HBs (e.g., Buonanno 1993; Fusi Pecci et al. 1993; 
Buonanno et al. 1997; Testa et al. 2001). 

For this reason, in the present study we decided to attempt a more 
comprehensive match between models and observations. Besides 
$B/(B+V+R)$, $V/(B+V+R)$, and $R/(B+V+R)$, we 
have also studied the behavior of the following additional HB morphology 
parameters:

\begin{itemize}
\item $B2'/(B+V+R)$: $B2'$ is defined as the number of 
HB stars {\em redder} than $(V-I)_0 = -0.02$~mag, and should not be 
confused with Buonanno's (1993) $B2$ parameter which represents 
the number of {\em blue} HB stars {\em bluer} than $(\bv)_0 = -0.02$~mag. 
In the case of NGC~362, we 
find a value $B2'/(B+V+R) = 0.99$ for NGC~362; 

\item $B0'/B$: $B0'$ is defined as the number of {\em blue} 
HB stars {\em redder} than $(V-I)_0 = +0.0$~mag.
We find $B0'/B \simeq 0.86$ for NGC~362; 

\item $\langle (V-I)_{\rm 0,RHB}\rangle$: This is simply the mean $(V-I)_0$ 
color of the red HB stars. 
We find $\langle (V-I)_{\rm 0,RHB} \rangle = +0.74$~mag for NGC~362; 

\item $\Delta (V-I)_{\rm RGB,RHB}$: This is defined as the difference 
in mean $V-I$ color between the red HB and the RGB at the same level 
(in $V$) as the red HB. In the case of NGC~362, we find a value  
$\Delta (V-I)_{\rm RGB,RHB} = 0.19$~mag.

\end{itemize}

From the data presented in Paper~I for NGC~362, one also finds 
$B/(B+V+R) = 0.06$, $V/(B+V+R) = 0.06$, and $R/(B+V+R) = 0.88$. 
To arrive at these figures, we had to 
deal with a number of (apparently) non-variable stars located in the 
instability strip region of NGC~362 and for which there is no 
membership information from the Tucholke (1992) proper-motion survey. 
As an objective criterion to assign these non-variable stars to either 
the red or blue HB components, we used the boundary between 
fundamental (RRab) and first-harmonic (RRc) pulsators for M3 
(NGC~5272) as given by Bakos \& Jurcsik (2000), namely: 
$(V-I)_0 \simeq 0.41$~mag. Any non-variable HB stars redder than this 
limit were assumed to be red HB stars, and vice-versa for the blue HB. 
(The number of such stars is quite small, as can be seen from Fig.~1,
so that our results are not seriously affected by the adopted procedure.)
The total number of HB stars, after statistically rejecting 10 red HB 
stars as possible non-members, is $B+V+R = 115$ for NGC~362.

Once these parameters had been obtained for NGC~362, 
extensive grids of synthetic HBs were computed 
aiming at estimating, by comparison between the observed and 
predicted HB morphology parameters, the best-fitting values of  
$\langle M_{\rm HB} \rangle$ (mean mass) and $\sigma_M$ (mass 
dispersion). We carried out computations for two difference 
metallicities, namely $Z = 0.001$ (corresponding to 
${\rm [Fe/H]} \simeq -1.5$ for $[\alpha/{\rm Fe}] = +0.3$) and 
$Z = 0.002$ (corresponding to 
${\rm [Fe/H]} \simeq -1.2$ for the same $[\alpha/{\rm Fe}]$ value; 
cf. Salaris, Chieffi, \& Straniero 1993). These two cases could 
be described as spanning the range in chemical abundances 
appropriate to the two clusters, particularly in view of the 
uncertainties involved in absolute measurements of [Fe/H] 
(i.e., the Zinn \& West 1984 vs. Carretta \& Gratton 1997 
metallicity scale controversy). We note, however, that the 
traditionally employed $Z = 0.001$ case appears likely to 
represent just a conservative lower limit to the metallicity 
of NGC~288 and NGC~362, even in the Zinn \& West scale.

\vskip 0.15in
\subsubsection{Best-Fitting Simulations: $Z = 0.001$ Case}

After inspection of several hundred sets of CMD simulations, 
we adopt the following parameters as providing the best  
global fit for NGC~362 in the $Z = 0.001$ case, in the canonical 
scenario: 

\vskip 0.125in
\begin{displaymath}
      \langle M_{\rm HB} \rangle = 0.706 \pm 0.005\, M_{\sun},\,\,\,
                       \sigma_M = 0.040 \pm 0.005\, M_{\sun}. 
\end{displaymath}
\vskip 0.125in

\noindent Three randomly picked CMD simulations (out of a pool of 
1000) for this best-fitting 
case are provided in Fig.~1 (left column, mid and bottom rows). 
ZAHB and TAHB (``terminal-age HB'') loci 
are overplotted in each panel as a reference. On the upper left 
panel of Fig.~1, we show the observed CMD from 
Paper~I (dereddened and shifted in magnitude by the indicated 
amounts) to enable direct comparison with the models. 

The error bars given above were estimated from extensive sets of 
simulations such as those summarized in Fig.~2 and Fig.~3. These two 
figures contain multiple panels that describe the dependence of the 
observables described 
in \S2.1.1 upon $\langle M_{\rm HB} \rangle$ and $\sigma_M$, 
respectively. To obtain Fig.~2, we held $\sigma_M$ fixed at the 
best-fitting value, and then allowed $\langle M_{\rm HB} \rangle$
to vary---and vice-versa for Fig.~3. In the 
top panel of both Fig.~2 and Fig.~3, the color difference between 
the RGB and the red HB is given; in the next panel, the mean color 
of the red HB is provided; in the third panel from the top, 
both $B2'$ and $B0'$ are shown; in the bottom 
panel, the behavior of the usual $B$, $V$, $R$ number counts is  
displayed. In all panels, 
the observed value for NGC~362 is indicated by a horizontal line.

\vskip 0.15in
\subsubsection{Best-Fitting Simulations: $Z = 0.002$ Case}
Following the same procedure as in the previous subsection---whose 
details we omit for conciseness---but 
assuming a higher metallicity for NGC~362 ($Z = 0.002$), we arrive
at the following parameters for the best-fitting Gaussian mass 
deviate: 

\vskip 0.125in
\begin{displaymath}
      \langle M_{\rm HB} \rangle = 0.667 \pm 0.005\, M_{\sun},\,\,\,
                       \sigma_M = 0.032 \pm 0.005\, M_{\sun}. 
\end{displaymath}
\vskip 0.125in

\noindent 
We note, at this point, that the ``color HB morphology parameters'' 
were not used to rule out any single 
($\langle M_{\rm HB} \rangle$,~$\sigma_M$) combination; rather, 
their inclusion in the analysis was intended primarily as a check of 
whether the colors of the models were in reasonable agreement with 
the observations. 
Note, in particular, that stars that arrive at the 
RGB phase with a higher mass are expected to be bluer than lower-mass 
giants, implying that $\Delta (V-I)_{\rm RGB,RHB}$ depends on the 
assumed RGB tip age. The $\Delta (V-I)_{\rm RGB,RHB}$ values reported 
in Fig.~2 and Fig.~3 were obtained from RGB 
tracks which imply an RGB tip age of about 14.5~Gyr. From the 
VandenBerg et al. (2000) models, one estimates a dependence of RGB 
color on age going as 
${\rm d} (V-I)_{\rm RGB}/{\rm d} t \approx 0.013$~mag/Gyr. This 
implies that the computed $\Delta (V-I)_{\rm RGB,RHB}$ values, as 
given in Fig.~2 and Fig.~3, would be in better agreement with the 
observed values in case the assumed age were lowered by 
$\sim 2$~Gyr---which still gives a reasonable absolute age for 
NGC~362. 
All in all, the colors predicted by the present 
simulations appear to be in good agreement with the observations 
at the $\approx 0.01$~mag level, both for NGC~288 and NGC~362.

\vskip 0.15in
\subsection{The Case of NGC~288}

\subsubsection{New HB Morphology Parameters: Definition}

As was the case with NGC~362 (\S2.1), we have computed extensive CMD 
simulations aiming at matching the observed HB morphology of NGC~288. 
Besides $B/(B+V+R)$, $V/(B+V+R)$, $R/(B+V+R)$, and 
$B2\arcmin/(B+V+R)$, we have also made an effort to reproduce the 
following additional HB morphology parameters:

\begin{itemize}
\item $\Delta V_{\rm tail}$: We compute the mean $V$ magnitude 
of the 10\% brightest blue HB stars (i.e., those close to the 
``horizontal'' level of the HB) and subtract this from the mean $V$ 
magnitude of the 10\% faintest blue HB stars (i.e., those located 
on the blue tail extension of the HB)---the resulting quantity 
is $\Delta V_{\rm tail}$. This is obviously  
intended to provide a quantitative measurement of the length of the 
blue HB tail. 
In the case of NGC~288, we find a value  
$\Delta V_{\rm tail} = 1.84$~mag, not taking into account the three 
possible ``extreme'' HB (EHB) stars discussed in Paper~I 
and which are much fainter than the bulk of the cluster's blue HB 
population; 

\item $B7'/(B+V+R)$: $B7'$ 
represents the number of {\em blue} HB stars redder than 
$(V-I)_0 = +0.07$~mag. This HB morphology indicator was chosen because,  
from the comparison between theoretical ZAHB models and the unreddened 
NGC~288 HB distribution, there appears to be a predominance of 
brighter-than-expected blue HB stars at the red end of the blue HB 
(see Fig.~4, top row; VandenBerg 2000, particularly his Fig.~28a).  We 
find $B7'/(B+V+R) = 0.13$ for NGC~288;  

\item $\langle (V-I)_{\rm 0,BHB}\rangle $: This is simply the mean $V-I$ 
color of the blue HB stars. Neglecting the three candidate EHB stars 
discussed in Paper~I,  
we find $\langle (V-I)_{\rm 0,BHB} \rangle = -0.03$~mag for NGC~288. 

\end{itemize}

Note also that, from the data presented in Paper~I, the traditional 
number count parameters $B/(B+V+R)$, $V/(B+V+R)$, and $R/(B+V+R)$ are 
found to be 0.98, 0.02, and 0, respectively. Utilizing the 
above reddening value we find, after ignoring the 
aforementioned EHB candidates, a value $B2'/(B+V+R) = 0.35$ for NGC~288. 

The total number of HB 
stars in NGC~288, including the 3 candidate EHB stars, is $B+V+R = 96$. 
In the simulations, this number was adopted, but the total number of 
stars in each synthetic HB model was allowed to fluctuate according to 
the Poisson distribution. 

Once these parameters had been obtained, extensive grids 
of synthetic HBs were computed in order to estimate the best-fitting 
$\langle M_{\rm HB} \rangle$ and $\sigma_M$ for NGC~288. As in the case 
of NGC~362, we carried out computations for two different metallicities, 
namely $Z = 0.001$ and $Z = 0.002$. As already noted, these two cases 
may be roughly considered representative of the Zinn \& West (1984) and 
Carretta \& Gratton (1997) scales, respectively.

%
\begin{figure*}[ht]
 \figurenum{4}
 \centerline{\epsfig{file=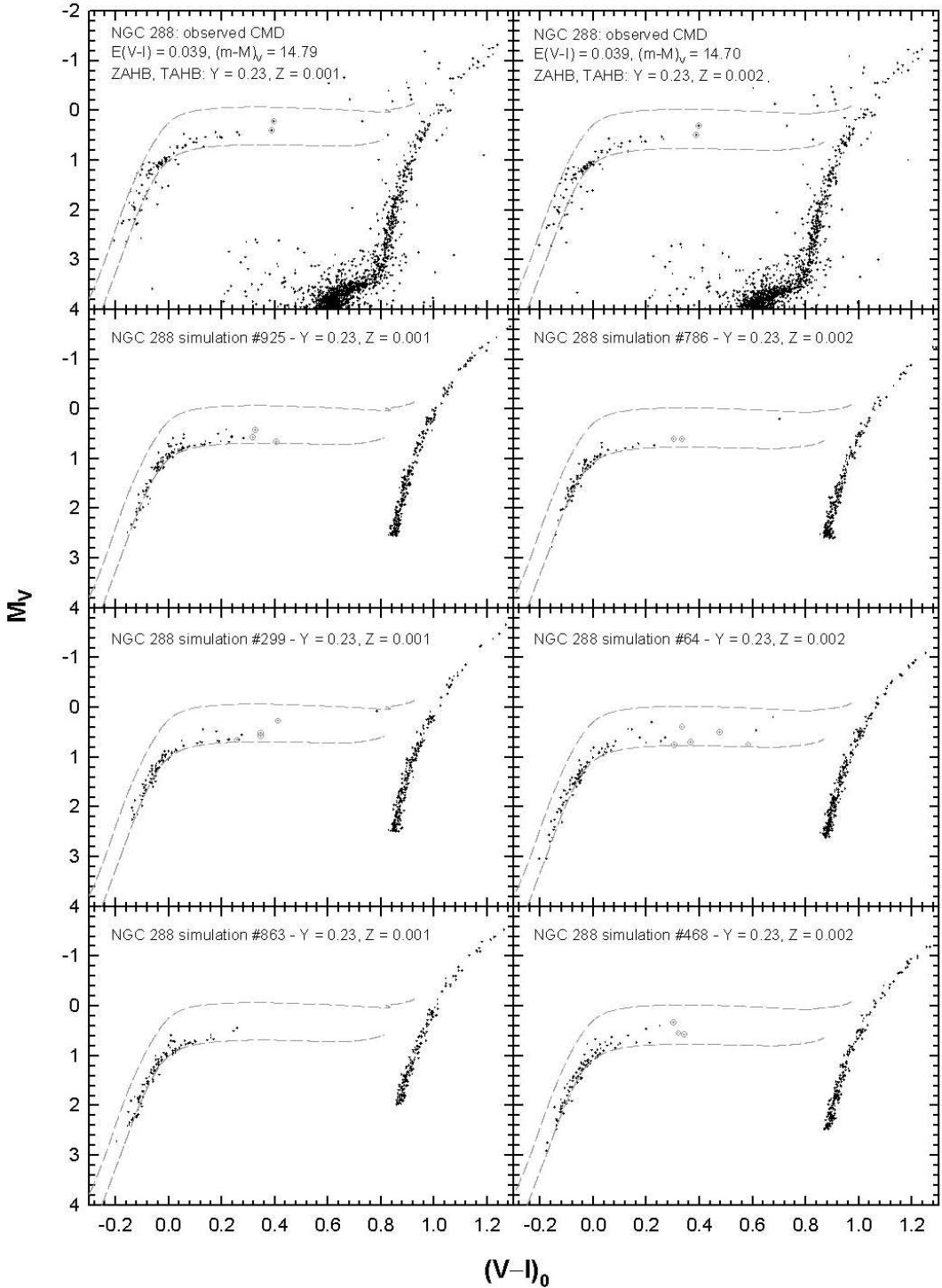,height=8in,width=5.75in}}
 \epsscale{0.5}
 \caption{As in Fig.~1, but for NGC~288. In this case, the reddening 
          value indicated, which is in excellent agreement with 
          Harris (1996) and Ferraro et al. (1999), was inferred by 
          assuming NGC~288 to have the same distance modulus as 
          derived for NGC~362 in Fig.~1. (The identity in distance 
          moduli between NGC~288 and NGC~362 follows from Paper~I.) 
         }
\end{figure*}

%
\begin{figure*}[ht]
 \figurenum{5}
 \centerline{\epsfig{file=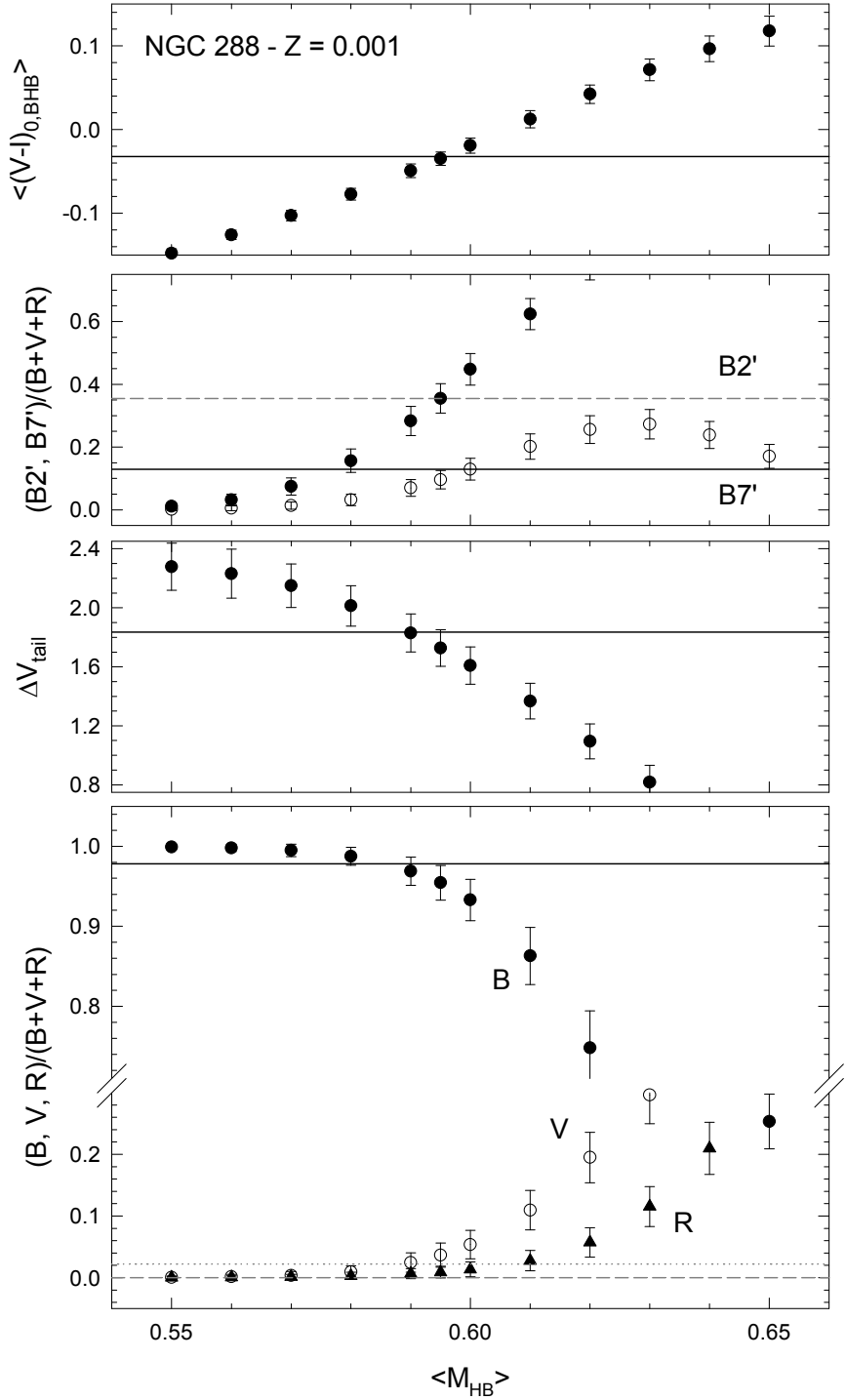,height=7.5in,width=4.5in}}
 \caption{As in Fig.~2, but for NGC~288. 
          Each datapoint was obtained from an average of 1000 HB 
          simulations. Horizontal lines indicate the observed 
          values for NGC~288. In the bottom panel, solid lines 
          indicate $B/(B+V+R)$, whereas dotted and dashed gray 
          lines denote $V/(B+V+R)$ and $R/(B+V+R)$, respectively. 
          In the second panel from the top, the solid line indicates 
          $B7\arcmin/(B+V+R)$, while the dashed gray line denotes 
          $B2\arcmin/(B+V+R)$.        }
\end{figure*}

%
\begin{figure*}[ht]
 \figurenum{6}
 \centerline{\epsfig{file=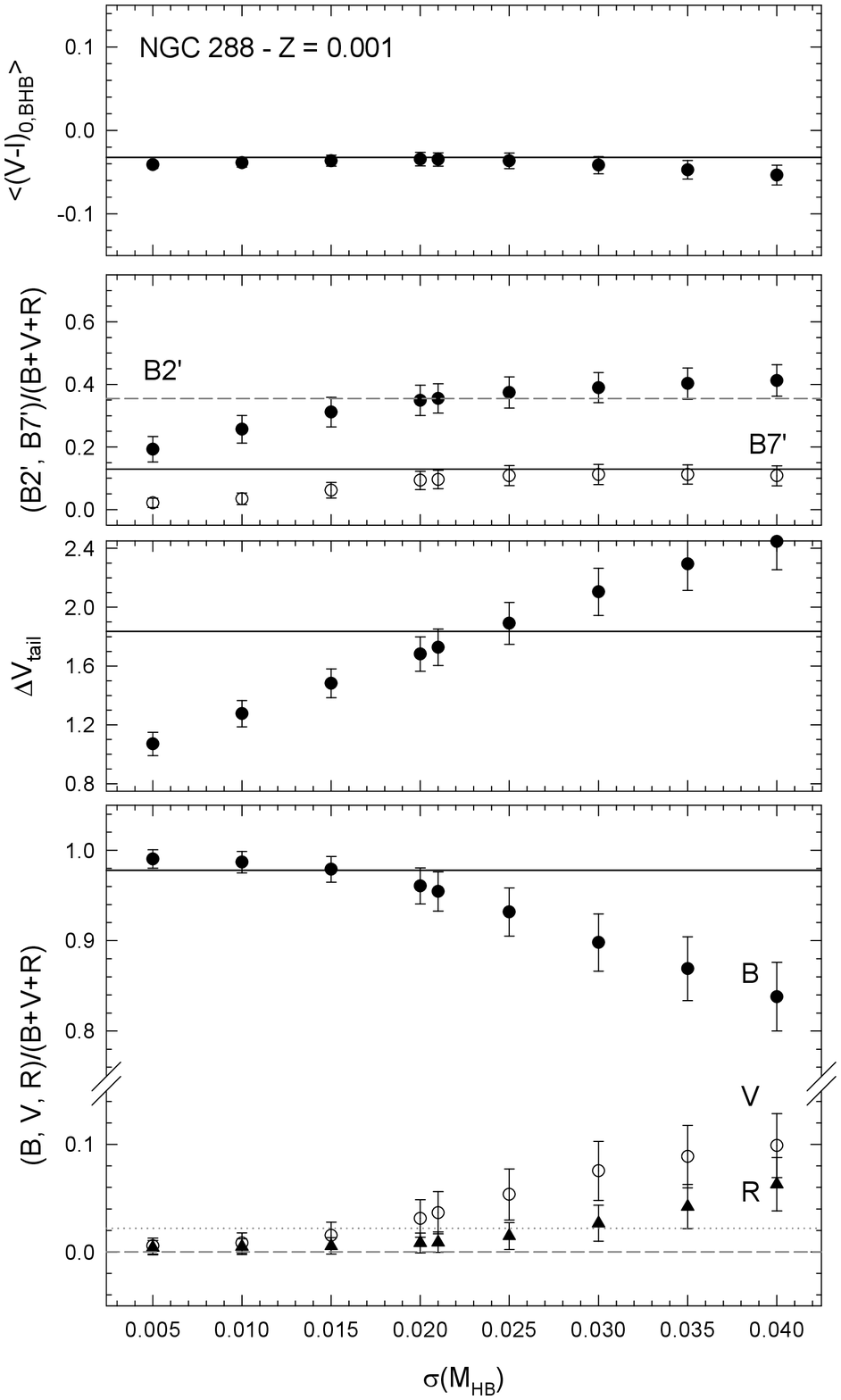,height=7.5in,width=4.5in}}
 \caption{As in Fig.~5, but for variations in $\sigma_M$.  
         }
\end{figure*}

\vskip 0.15in
\subsubsection{Best-Fitting Simulations: $Z = 0.001$ Case}
Proceeding in the same manner already described for NGC~362 in 
\S2.1, we arrived at the following best-fitting parameters for 
NGC~288, in the more metal-poor case: 

\vskip 0.125in
\begin{displaymath}
      \langle M_{\rm HB} \rangle = 0.595 \pm 0.005\, M_{\sun},\,\,\,
                       \sigma_M = 0.021 \pm 0.005\, M_{\sun}. 
\end{displaymath}
\vskip 0.125in

\noindent Three randomly picked CMD simulations (out of a pool of 
1000) for this best-fitting case are provided in the middle and 
bottom left panels of Fig.~4, which is analogous to Fig.~1 for 
NGC~288. Again in analogy with the analysis 
presented in \S2.1 for NGC~362, the error bars in the derived values 
of $\langle M_{\rm HB}\rangle$ and $\sigma_M$ for NGC~288 were 
obtained from extensive sets of panels similar to those depicted 
in Fig.~5 and Fig.~6, respectively. It is worth noting that the 
newly introduced 
$\Delta V_{\rm tail}$ parameter strongly constrains both the mean 
ZAHB mass and the dispersion around this mean value.

It is interesting to check these simulation results in more detail 
in order to determine whether the candidate EHB stars discussed in 
Bellazzini \& Messineo (1999) and in Paper~I may be 
a ``natural'' occurrence in this type of model. Among the pool of 
1000 simulations 
for the best-fitting ($\langle M_{\rm HB} \rangle$,~$\sigma_M$) 
combination, we find that only seven synthetic HBs contain EHB 
stars---where we define as an EHB star any star hotter than 
$T_{\rm eff} = 20,\!000$~K. Moreover, only one EHB star was 
present in each of these seven synthetic CMDs.

\vskip 0.15in
\subsubsection{Best-Fitting Simulations: $Z = 0.002$ Case}
Following the same procedure as in the previous subsection, but 
adopting a higher metallicity for NGC~288 ($Z = 0.002$), we arrive
at the following best-fitting (canonical) solution: 

\vskip 0.125in
\begin{displaymath}
      \langle M_{\rm HB} \rangle = 0.575 \pm 0.005\, M_{\sun},\,\,\,
                       \sigma_M = 0.018 \pm 0.005\, M_{\sun}. 
\end{displaymath}
\vskip 0.125in

A random sample of three simulations for this best-fitting combination 
of parameters is also displayed in Fig.~4 (right column, middle 
and bottom panels). The upper regions of the observed CMD of NGC~288 
(from Paper~I) are also given in the top right panel; the distance 
modulus employed was inferred in the same way as previously described 
for $Z = 0.001$. 
The ZAHB and TAHB loci for $Z = 0.002$ are overplotted. 
The reported errors are based on analysis entirely analogous to 
that described in the previous subsection for the $Z = 0.001$ case. 

As far as EHB stars are concerned, we find that the situation does 
not change much with respect to the results obtained in the previous 
subsection for $Z = 0.001$. In particular, among the pool of 
1000 simulations for the best-fitting 
($\langle M_{\rm HB} \rangle$,~$\sigma_M$) combination, only ten CMD 
simulations contain EHB stars (again with a single EHB star present 
per model). It thus appears that if the three candidate EHB stars in 
NGC~288 turn out to be bona-fide cluster members, they must be 
explained by a process different from the one that 
generated the bulk of the HB population of the cluster---which 
is perhaps not surprising, particularly in view of the fact that these 
EHB stars are much fainter than the bulk of the blue HB population of 
the cluster.  

To close this section, we note that the dispersion in HB mass that 
we find for NGC~288 is clearly smaller than for NGC~362. We suggest 
that this difference may be related to the higher core density in 
NGC~362. Interestingly, Catelan et al. (2001) found that the loose, 
outer-halo globular cluster Palomar~3 is characterized by an even 
smaller dispersion in HB mass than we found here for NGC~288. 
We will come back to this subject in \S5.

\vskip 0.15in
\subsubsection{``Anomalous'' Red Blue HB (ARBHB) Stars}
Figure~4 reveals a rather satisfactory 
agreement between the models and the observations of NGC~288.  
However, there is a noticeable discrepancy at the redder part of the 
blue HB [$0.05 \lesssim (V-I)_0 \lesssim 0.3$] where the observed 
number of stars brighter than the
ZAHB is significantly larger than in the models (see also VandenBerg 2000).
Hereafter, we shall call these ``anomalous red blue 
HB'' (ARBHB) stars. The extent to which these stars are 
brighter than the ZAHB depends somewhat on the adopted reddening, 
as can also be seen from Figs.~28a,b of VandenBerg (2000).\footnote{
Note that reddening uncertainties also introduce some leeway 
in the inferred difference in distance moduli 
($\Delta\mu$, following the notation of Paper~I) between NGC~288 
and NGC~362 derived from fits of the data to theoretical ZAHBs:
Experiments utilizing the reddening values from Harris (1996),  
Schlegel et al. (1998) and Ferraro et al. (1999) clearly indicate 
that the $\Delta\mu$ value obtained by the ZAHB-fitting procedure is 
uncertain by $\pm 0.1$~mag. We emphasize that the purpose of the 
present study is {\em not} to derive $\Delta\mu$ through ZAHB-fitting, 
but instead to analyze the ``horizontal'' distribution of stars along 
the HBs of NGC~288 and NGC~362.} 
As far as the apparent overluminosity of these ARBHB stars is concerned, 
we suggest that the best approach to reliably study this phenomenon 
is through analysis of spectroscopic gravities and ultraviolet 
photometry---$B,V,I$ photometry providing an interesting first 
indication that a problem may exist, but being clearly inadequate  
to examine the phenomenon in sufficient detail. 

In fact, there is some 
evidence that ARBHB stars may be present in several other 
globular clusters (see also Markov, Spassova, \& Baev 2001). 
This is suggested, in particular, by the $u$, 
$(u-y)_0$ diagrams for blue-HB clusters in Fig.~1 of Grundahl et 
al. (1999): In the cases of NGC~288, M12 (NGC~6218), M13 (NGC~6205), 
NGC~6752, NGC~6397, and M56 (NGC~6779), 
the redder of the blue HB stars seem brighter than the ZAHB 
at the observed colors. Lee \& Carney (1999, see Appendix~A in 
their paper) have discussed what appears to be another instance of 
the same effect, this time at the RR Lyrae level of 
the blue-HB cluster M2 (NGC~7089). In this sense, 
it is possible that the ARBHB star phenomenon is related to 
the  long-standing difficulties in accounting 
for the large number of  ({\em presumably} ``evolved'') RR Lyrae stars 
in metal-poor,  Oosterhoff type II globular clusters with 
intermediate-blue HBs, such  as M15 (NGC~7078; see, e.g., 
Renzini \& Fusi Pecci 1988; Rood \& Crocker 1989). 
Thus,  a resolution of the ARBHB stars problem may 
have (potentially) far-reaching implications. 

At least in a qualitative sense, some of these ``evolved'' stars 
might be (partly) accounted for if the mass distribution were 
actually bimodal (or if the ZAHB mass distribution contained a 
cutoff at the red end; see also Appendix~A in Lee \& Carney 1999) 
and/or if the duration of the late stages 
of the evolution of HB stars has been somewhat underestimated in 
extant evolutionary computations. One possible cause for the latter 
could be the very uncertain $^{12}{\rm C}(\alpha,\gamma)^{16}{\rm O}$ 
reaction rate, which constitutes the dominant source of energy  
towards the end of HB evolution (e.g., Imbriani et al. 2001 and  
references therein). However, Renzini \& Fusi Pecci 
(1988) and Dorman (1992) have pointed out that a change in 
the rates of this reaction by a factor of three would change the 
HB lifetime by only about 10\% (see also Brocato, Castellani, \& 
Villante 1998). 
Another well-known source of uncertainty in the 
computed duration of the HB evolutionary phase is the treatment 
of semiconvection and ``breathing pulses'' (e.g., Bressan, Bertelli, 
\& Chiosi 1986; Caloi \& Mazzitelli 1993; Imbriani et al. 2001).  
A detailed analysis of the problem of ARBHB stars is 
beyond the scope of the present paper and will be deferred to a 
future occasion.

It is important to note, in any case, that the number of ARBHB 
stars in NGC~288 is relatively small---less than one fifth of the 
cluster's HB population---so that the main results of our analysis 
should not be substantially affected by our difficulties in 
accounting for these stars, particularly if a solution to the 
problem can be found within the canonical framework. 
On the other hand, our results {\em could} be 
significantly affected, if the ARBHB stars point to some 
non-canonical effect affecting {\em the bulk} of NGC~288's blue 
HB stars. This is obviously a consequence of the fact that the  
present simulations are entirely based on {\em canonical} models.

\vskip 0.15in
\section{Estimating Relative Ages from Differences in the HB Morphology}

In order to estimate the differential ages required to produce the
relative HB types of NGC~288 and NGC~362, we follow the same approach
described in Catelan (2000) and Catelan et al. (2001) in the cases of
M5 vs. Palomar~4/Eridanus and M3 vs. Palomar~3, respectively. 
As in those 
papers, we shall evaluate the effects of an age-dependent mass loss 
on the RGB, as implied by the several different analytical formulae 
discussed in the Appendix to Catelan's paper. 
As discussed by Catelan, at the present 
time it does not appear possible to strongly argue in favor of any of 
his recalibrated equations over the others, so that a safer approach 
is to use all of them simultaneously whenever an estimate of the amount
of mass loss on the RGB is needed. We briefly recall that equations~(A1) 
through (A4) of Catelan are ``generalized,'' empirically recalibrated 
variations of the analytical mass loss formulae previously suggested 
by Reimers (1975a, 1975b), Mullan (1978), Goldberg (1979), and Judge 
\& Stencel (1991), respectively.

In this paper, RGB mass loss was estimated on the basis of the
RGB models of VandenBerg et al. (2000) for a chemical composition
${\rm [Fe/H]} = -1.54$, $[\alpha/{\rm Fe}] = +0.3$ ($Z = 0.001$ 
case) and ${\rm [Fe/H]} = -1.31$, $[\alpha/{\rm Fe}] = +0.3$ 
($Z = 0.002$ case). 

Using these ingredients, HB morphology-based relative ages (sense NGC~288 
minus NGC~362) were computed according to the Catelan (2000) prescriptions, 
and are shown graphically in Fig.~7. 
This figure gives the age difference (in 
Gyr) between NGC~288 and NGC~362 as a function of the NGC~288 absolute 
age (also given in Gyr) 
for the several indicated analytical mass loss formulae (from the Appendix 
in Catelan's paper). The upper panel refers to the $Z = 0.001$ case, and 
the lower panel to the $Z = 0.002$ case. In this figure, the shaded 
regions denote the TO age difference, $\Delta t \approx 2\pm 1$~Gyr, 
estimated in Paper~I using the ``bridge test.''\footnote{It is well known that, 
at a given chemical composition and difference in TO luminosity, 
the age difference between two clusters depends on the absolute age of the 
oldest cluster: Larger absolute ages correspond to larger age differences 
(see Paper~I). For the same reason a given error bar in the observable
``difference in TO luminosity'' corresponds to larger error bars in ``age
difference'' for large absolute ages and smaller ones for low absolute ages.
The shaded area of Fig.~7 fully accounts for these effects. Its boundaries have
been obtained by estimating the age difference between NGC~288 and NGC~362
and the associated error bars with the procedure adopted in Paper~I 
(see Fig.~11 and Fig.~12 there), while varying the assumptions on the absolute age
of NGC~288. Note that the best fitting absolute age for NGC~288 derived in 
Paper~I is $\sim 13-14$~Gyr. 
} 

Selected results are given, in the case of Reimers' (1975a, 1975b) formula, 
in Table~1. (The results are qualitatively similar for the other employed 
mass loss formulae, and are omitted for conciseness.) 
Besides the absolute HB morphology ages for NGC~288 and 
NGC~362 (columns 2 and 3; the implied age differences are given in column 8), 
the corresponding RGB tip masses when mass loss is ignored 
($M_{\rm RGB}^{\rm tip}$) 
and overall amount of mass loss on the RGB ($\Delta M$), both in solar units, 
are also given (columns 4,5 and 6,7, respectively). For reference purposes, 
the implied 
HB morphology age of the ``inner halo globular clusters'' (IHGCs) isochrone 
(see Catelan \& de Freitas Pacheco 1993) at the same metallicity is given 
in the first column. 

In the next section, we discuss the important implications of Fig.~7  
for our understanding of the second parameter effect.

\vskip 0.15in
\section{Age as the Second Parameter in NGC~288/NGC~362?} 

As reviewed in \S1, previous theoretical analyses of the HB morphologies  
of NGC~288 and NGC~362 have required fairly large age differences, 
generally in 
excess of 3~Gyr, to account for this second parameter pair entirely in 
terms of age. This is not the case in the present study: For the first 
time, we are clearly successful at providing a satisfactory description 
of the NGC~288/NGC~362 pair under the assumption that age is the sole 
second parameter. 

However, this conclusion does depend on certain conditions. As one can 
see from Fig.~7 (upper panel), if we assume a ``canonical'' metallicity 
$Z \simeq 0.001$ for this pair, as has been done in all previous 
theoretical analyses (Lee et al. 1988, 1994; Catelan \& 
de Freitas Pacheco 1994), one still finds some difficulty in accounting 
for the observed second parameter effect entirely in terms of age. 
In particular, relative ages smaller than 2~Gyr do seem to require an 
absolute NGC~288 age smaller than $\approx 10$~Gyr. As indicated in 
Fig.~7, age differences smaller than 2~Gyr are certainly within the 
allowed range of the study presented in Paper~I; moreover, as we have 
previously discussed (\S1), some authors still favor an age difference 
of 1~Gyr (or less) for this pair. Whether absolute ages as low as about 
10~Gyr are acceptable is, of course, a matter of serious debate (see, 
e.g., Grundahl et al. 2000 for a recent discussion).

%
\begin{figure*}[ht]
 \figurenum{7}
 \centerline{\psfig{figure=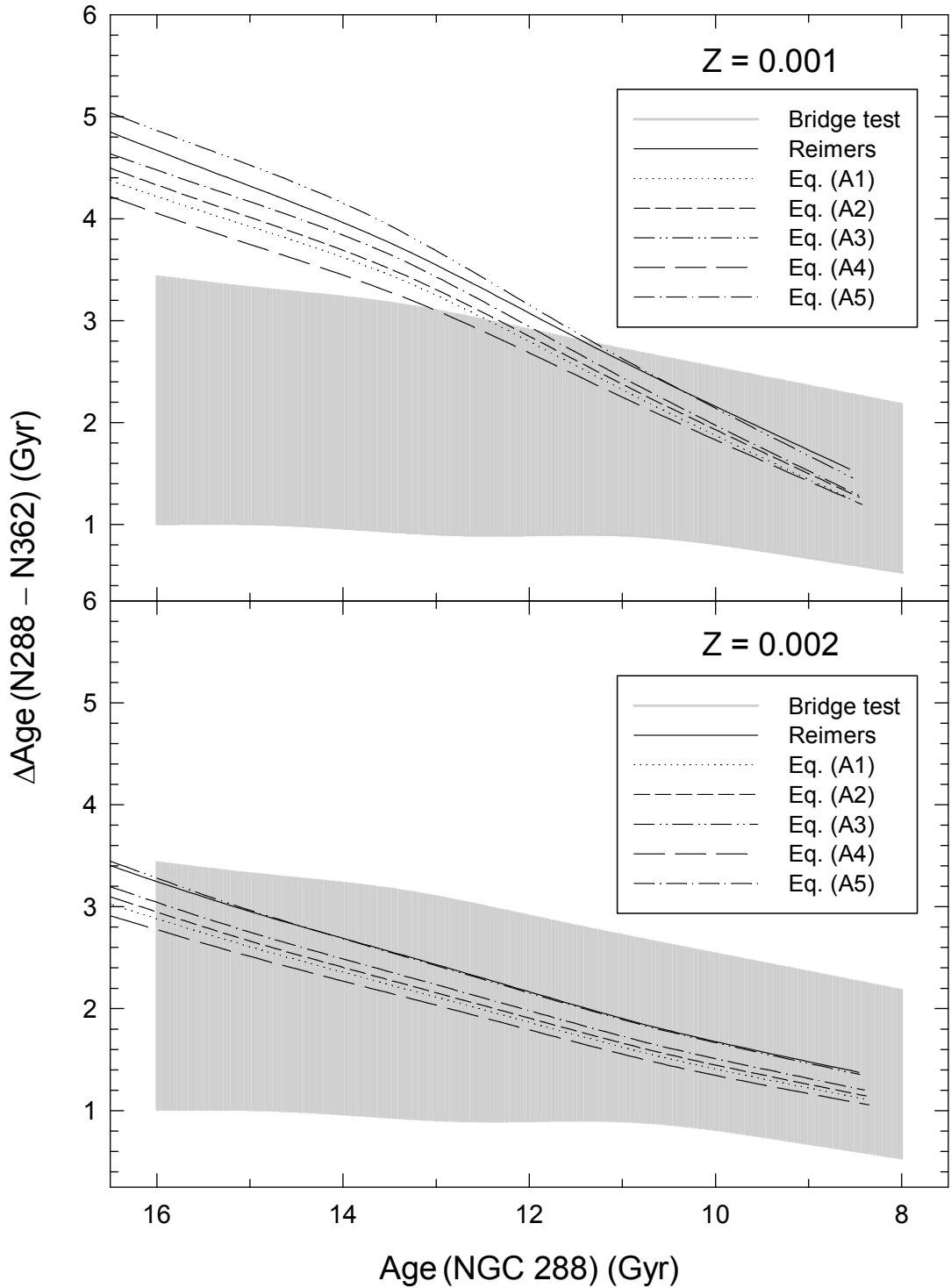}}
 \caption{The ``HB morphology-based'' difference in age between 
    NGC~288 and NGC~362 (given 
    in Gyr), as derived for the several indicated mass loss formulae 
    for red giant stars
    (cf. Catelan 2000), is plotted as a function of the absolute NGC~288 
    age for two different assumptions on the metallicity: $Z = 0.001$ 
    (upper panel) and $Z = 0.002$ (bottom panel). The shaded 
    band indicates the TO age difference range favored 
    by our application of the ``bridge method'' (Paper~I) as a function
    of absolute age (see \S3). Its boundaries 
    have been obtained by estimating the age difference between 
    NGC~288 and NGC~362 and the associated error bars with the procedure 
    adopted in Paper I (see Fig.~11 and Fig.~12 there), while varying the 
    assumptions on the absolute age of NGC~288. Note that the best fitting 
    absolute age for NGC~288 derived in Paper~I is $\sim 13-14$ Gyr. 
  }
\end{figure*}

\begin{figure*}[ht]
 \centerline{\epsfig{file=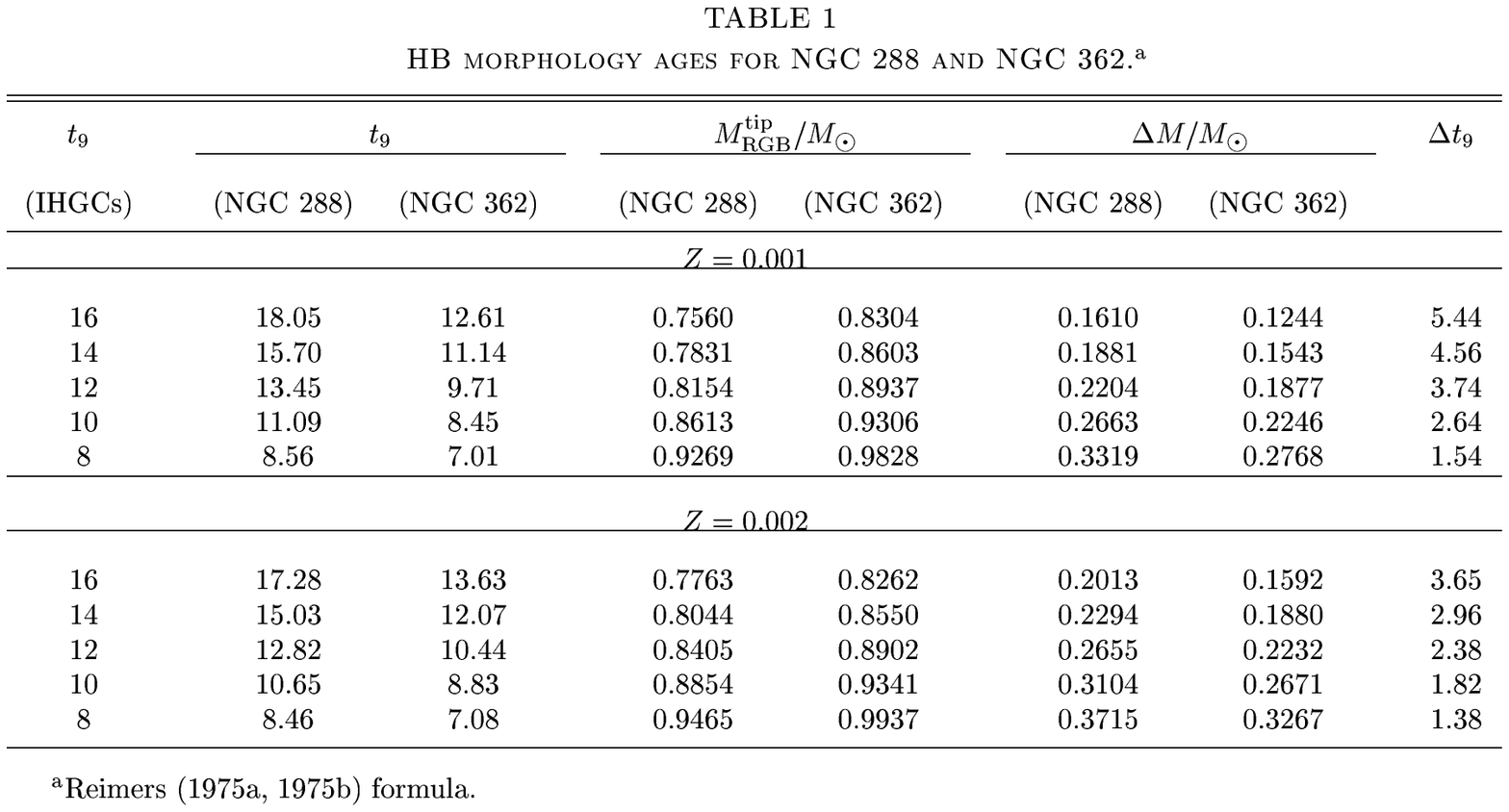,height=2.85in,width=5.35in}}
\end{figure*}

It is only when a higher metallicity is assumed for the NGC~288/NGC~362 
pair that one can seriously argue that a solution to the second parameter 
problem, in the case of this classical second-parameter pair at least, 
may have been found. It is quite evident from Fig.~7 (bottom panel) 
that, for a metallicity $Z = 0.002$, turnoff age differences as high as 
3~Gyr can comfortably account for the difference in HB types between 
NGC~288 and NGC~362: Absolute NGC~288 ages as high as 16~Gyr are still 
consistent with the ``age as the second parameter'' hypothesis. 
Importantly, one also finds that age differences in the $1.5-2$~Gyr 
range can be accommodated, for such a higher $Z$, assuming absolute 
NGC~288 ages in the $9-13$~Gyr range. 

Again, we emphasize that this 
result was not present in previous investigations simply because a 
substantially lower metal abundance had been adopted in all such 
analyses. On the other hand, it had been anticipated many times 
in the literature that an increase in the metallicity would lead
to a larger sensitivity of the temperature of HB stars to small changes in
mass (see, for example, Buonanno, Corsi, \& Fusi Pecci 1985, in particular 
their Fig.~6), thus alleviating the second parameter problem.

Analysis of Table~1 discloses that, while the effect of metallicity 
on RGB mass loss also contributes to decreasing the HB morphology age
difference for a given absolute age, the chief cause of a decrease in 
such relative ages is the difference in HB mass derived from the 
synthetic HBs (\S2).   

Since a higher metallicity appears to be highly favorable to the 
hypothesis that age is the second parameter of HB morphology for the 
NGC~288/NGC~362 pair, it is clearly of the utmost importance that the 
existing discrepancies between the Zinn \& West (1984) and Carretta 
\& Gratton (1997) metallicity scales be resolved. As well known, at 
such relatively high metallicities, the latter scale favors 
significantly higher [Fe/H] values than found in the more traditional 
Zinn \& West scale; it thus follows that the Carretta \& Gratton 
scale is also more favorable to the ``age as the second parameter'' 
hypothesis than is the Zinn \& West scale. Interestingly, however, 
the possibility that the high-resolution spectroscopy scale of 
Carretta \& Gratton may overestimate the globular cluster 
metallicity, and perhaps be more similar to the Zinn \& West scale 
than had previously been realized, has recently been addressed by 
Bragaglia et al. (2001) and Frogel (2001). In any case, it is 
important to note that the $Z = 0.001$ case is likely to provide 
just a conservative lower limit to the actual metallicity of the 
pair NGC~288/NGC~362, even in the Zinn \& West scale. An equally 
conservative upper limit to the clusters' metallicities, in the 
Carretta \& Gratton scale, would clearly correspond to $Z > 0.002$.

It is instructive to compare our derived difference in HB mass between 
NGC~362 and NGC~288 with those from previous 
analyses of this pair. While Lee et al. (1988) report a difference in 
$\langle M_{\rm HB}\rangle$ of $0.125\, M_{\odot}$ between NGC~362 
and NGC~288, this value was later revised by Lee et al. (1994) to 
$0.094\, M_{\odot}$. Note that these authors assumed $Z = 0.001$ for 
NGC~362, but an even lower metallicity, $Z = 0.0007$, for NGC~288; had 
they assumed the same $Z$ for the two clusters, their derived HB mass 
difference would have been accordingly higher (compare the 
$\langle M_{\rm HB}\rangle$ values derived for NGC~288 in \S2.2.2 and 
\S2.2.3). Indeed, assuming the same $Z = 0.001$ for the two 
globulars, Catelan \& de Freitas Pacheco (1994) estimated a difference 
in mean HB mass of $0.126 - 0.139 \, M_{\odot}$. The present, more 
comprehensive HB morphology analysis, on the other hand, favors a 
smaller mass difference, namely: $0.111 \, M_{\odot}$ ($Z = 0.001$) 
or $0.092 \, M_{\odot}$ ($Z = 0.002$); but note that a slightly larger 
difference in mean HB mass (by $\simeq 0.005-0.01\, M_{\odot}$) 
{\em would} have been found 
had we, as usually done, relied solely on the traditional number 
count parameters $B$, $V$, $R$ (see the lower panel of Fig.~5). Such a 
decrease in the differential HB mass estimate, along with the fact that 
the mass loss on the RGB has been assumed, in line with Catelan's (2000) 
prescriptions, to increase with age, explains the different conclusions 
reached in this paper in comparison with previous ones.  

We emphasize that while age may be the explanation for the 
difference in HB populations of these two clusters, this explanation
is not without difficulty. In particular, we confirm a previous suggestion 
(VandenBerg 2000) that canonical models are unable to reproduce the 
detailed HB morphology of NGC~288 at its red end, giving rise to the 
``ARBHB'' stars, for as yet unknown 
reasons. There is evidence that ARBHB stars may be present in several 
other blue-HB clusters, and the phenomenon may also be related to 
well-known (but often overlooked) problems encountered in the 
modelling of the RR Lyrae variables in Oosterhoff type II 
globulars which do not have extremely blue HBs, such as M15 (see \S2.2.4). 
Whether this will simply require some tweaking of the current 
HB evolution ingredients (e.g., reaction rates, treatment of 
semiconvection and ``breathing pulses'') within the canonical framework, 
or instead require a much more drastic rupture of the canonical scenario, 
is still unclear.

Another problem that the current evolutionary models seem to face, at 
least in the cases of NGC~362 and NGC~1851, concerns the reproduction 
of the detailed morphology of the red clump. As can be seen by careful  
inspection of Fig.~1, the simulations show a gentle slope at the red 
end of the red HB which is not obvious in the data. Judging from 
the ZAHB fits presented in his Figs.~28c,d, the same problem 
is present also in the case of VandenBerg's (2000) analysis. 
One possible explanation, in analogy with Dorman, VandenBerg, \&  
Laskarides (1989), is that NGC~362 and NGC~1851 have a somewhat 
higher helium abundance 
than assumed in the present model computations---which employ the 
traditional $Y_{\rm MS} = 0.23$ value. 
We have utilized the Sweigart \& Catelan (1998) HB tracks 
for various helium abundances and a metallicity $Z = 0.002$ in an 
examination of this problem, and found that a $Y_{\rm MS} \simeq 0.28$ 
would provide a better match 
between the data and the models, as far as the slope of the red end 
of the red HB is concerned. Unfortunately, it is not possible to apply  
a similar test in the case of NGC~288, given its exclusively blue HB. 
A higher helium abundance in NGC~362 than in NGC~288 would go in the 
opposite sense to that needed 
to reproduce the difference in HB types between these clusters.  
While an application of Iben's (1968) ``$R$-method'' to the datasets 
presented in Paper~I does not disclose an obvious difference in the 
clusters' helium abundances---we find, following the Buzzoni et al. 
(1983) definition, $R = 1.27\pm 0.28$ for NGC~288, implying 
$Y_{\rm MS}^{\rm NGC~288} \approx 0.22 \pm 0.04$; 
and  $R = 1.10\pm 0.20$ for NGC~362, giving 
$Y_{\rm MS}^{\rm NGC~362} \approx 0.19 \pm 0.03$---the 
uncertainties in the helium abundance thus derived 
remain sufficiently large that neither the possibility of 
a higher $Y_{\rm MS}$, 
nor that of a difference in helium abundance, can be conclusively  
ruled out. HB simulations similar to those presented in \S2 for 
$Z = 0.002$ but assuming a higher helium abundance $Y_{\rm MS} = 0.28$ 
for both clusters give a 
difference in mean HB mass of $\approx 0.11\, M_{\odot}$ between 
NGC~362 and NGC~288---which 
is somewhat larger than the corresponding value ($0.092\, M_{\odot}$)
for $Y_{\rm MS} = 0.23$. Even though the mass difference for the 
higher-$Y_{\rm MS}$ case 
cannot be converted into an HB morphology-based age difference at this 
point (due to the lack of sufficiently extensive grids of RGB tracks 
and isochrones 
for non-canonical helium abundances), the present exercise 
should suffice to show that the uncertainty in the HB morphology-based 
age difference between NGC~288 and NGC~362 stemming from a possible 
systematic uncertainty in the helium abundance is of the same order of 
the uncertainty in the metallicity scale discussed previously.

In addition, it should be noted that the assumption of identical 
heavy-element abundances for NGC~288 and NGC~362 may not necessarily 
be valid. In particular, Shetrone \& Keane (2000) suggest that there 
may be 
ab initio differences, particularly in the Ca and O abundances, 
between these two clusters: NGC~288 stars may be overabundant in 
these elements by about 0.1--0.15~dex compared to NGC~362.\footnote{
Shetrone \& Keane (2000) do not include carbon or nitrogen abundances in their
analysis, but  the earlier results of Dickens et al. (1991)---unfortunately, for 
a significantly smaller sample of stars---suggest that there is no significant
variation of 
${\rm [(C+N+O)/H]}$ between NGC~288 and NGC~362.
In addition, the oxygen abundance derived from atomic lines may depend upon the
assumed carbon abundance, which determines the fraction of oxygen found in  
the CO molecule (e.g., Milone et al. 1992; Mel\'endez, Barbuy, \& Spite
2001).}
Note that the sense of the 
difference goes {\em against} that needed to explain the difference 
in HB morphology between these two clusters. On the other hand, there 
is also an indication that NGC~288 may be more deficient in iron than 
NGC~362, by about 0.06~dex---which would compensate (at least in part) 
the effects of the suggested Ca and O abundance differences upon the 
clusters' HB morphologies, and thereby on the inferred HB 
morphology-based age difference. 

It should also be noted, in line with Catelan (2000), that the 
treatment of mass loss on the RGB using analytical mass loss formulae, 
a procedure upon which the present results depend, may not be entirely 
appropriate (see also Willson 2000).  Our knowledge of mass loss in 
low-mass, low-metallicity 
red giants is---at best---at a rudimentary stage. One particularly 
tantalizing, if speculative, 
possibility is that the correlation between stellar rotation 
rates in blue HB stars and HB morphology also implies a correlation 
between RGB mass loss and rotation; this could help explain Peterson's 
(1985) observation that blue HB stars in NGC~288 rotate faster than in 
(more metal-poor) clusters with much redder HBs, such as M3, 
M4 (NGC~6121), and M5. On the other hand, Shetrone \& Keane 
(2000) have recently conducted a study of spectroscopic mass loss 
indicators in NGC~288 and NGC~362, and suggested that RGB stars 
in these two clusters have very similar mass loss rates. If true, this 
would rule out the correlation between stellar rotation and RGB mass loss 
rates  as a viable hypothesis to account for (part of) the second-parameter 
problem in the case of this pair. In our opinion, 
however, the Shetrone \& Keane results are far from being conclusive. 
In particular, their sample is still inadequate to study differences 
in mass loss rates between giants in the two clusters. This is evident 
from Fig.~5 in their paper, where one clearly finds that the vast 
majority of their brighter studied stars are members of NGC~362, whereas 
the NGC~288 sample is much fainter. Therefore, the regime where more 
extreme mass loss might be expected for NGC~288 was simply not covered 
in the Shetrone \& Keane investigation, and the question whether 
NGC~288 giants may lose more or less mass than NGC~362 giants close to 
the tip of the RGB remains open.\footnote{All analytical mass loss 
formulae that 
we are aware of, including those from Catelan (2000), imply that giants 
in an older globular cluster {\em should} lose more mass than giants in 
a younger globular with the same chemical composition. This is clearly
illustrated in Table~1.}  

Last, but not least, it should be noted that the 
cause of the bimodal HB of NGC~1851 is currently unknown, but 
seems unlikely to be due to an intrinsic age spread (see \S3.2 in 
Paper~I). Therefore, while 
age may be the second parameter in the case of NGC~288/NGC~362, an  
additional second parameter seems to be required to fully explain  
the HB morphologies of NGC~1851/NGC~288/NGC~362. Whether this will 
eventually require a change in the relative turnoff ages estimated 
in Paper~I or not remains unclear at the present time.

\vskip 0.15in
\section{HB Mass Dispersion--Central Density: A Correlation?}

Another important result of our analysis is that NGC~362's 
HB appears to have a much wider dispersion 
in mass than is the case for NGC~288. We propose that there is a 
correlation between the central density of globular clusters and mass 
dispersion on the HB, which is presumably related to the way a dense  
environment is able to affect mass loss on the RGB (Fusi Pecci et al. 
1993; Buonanno et al. 1997, 1999; Testa et al. 2001). 
Whereas NGC~362 is a very dense cluster 
($\rho_0 = 5\times 10^4\,L_{\odot}{\rm pc}^{-3}$; Harris 1996) with  
a possibly collapsed core (e.g., Trager, Djorgovski, \& King 1993), 
NGC~288 is very loose ($\rho_0 = 68\,L_{\odot}{\rm pc}^{-3}$, $c = 0.96$). 
Hence while it is unlikely that the NGC~288 blue HB population should
have its origin attributed to environmental effects, the same cannot be said 
with any degree of confidence about NGC~362. In other words, if NGC~362 
had the same structural parameters as NGC~288, one might plausibly expect 
it to have an even redder HB, perhaps completely devoid of blue HB and 
RR Lyrae stars. Another possible hint that the HB morphology of NGC~362 
is affected by its high central density are the 14 blue stars detected 
within the inner $14\arcsec$ of NGC~362 using ultraviolet imagery from 
UIT and WFPC2 (Dorman et al. 1997). Unfortunately, it is not yet known 
whether these are true HB stars or extreme blue stragglers (which are 
known to show a strong radial dependence). 

The same idea may be applicable to the case of 
NGC~1851, a cluster whose bimodal HB {\em may}  
be compatible with a unimodal HB mass distribution,  
provided the mass dispersion on the ZAHB is very large---i.e., 
substantially larger than found here for NGC~362 (Catelan et al. 1998). 
Indeed, NGC~1851 appears to have 
the same age as NGC~362 (Paper~I and references therein), 
but it also has a more sizeable fraction of blue HB and RR Lyrae stars. 
Importantly, NGC~1851 is even denser than NGC~362, with 
$\rho_0 = 2.1\times 10^5\,L_{\odot}{\rm pc}^{-3}$. Note also that, 
consistent with this trend, the extremely loose outer-halo 
cluster Palomar~3 has an even smaller dispersion in mass on the HB 
than NGC~288 (Catelan et al. 2001). A systematic study of   
a larger sample of globulars would be useful to determine the  
extent to which this interesting correlation holds. 

In conclusion, 
we note that, if central concentration is indeed {\em a} second parameter 
(Fusi Pecci et al. 1993; Buonanno et al. 1997, 1999; Testa et al. 2001), 
driving the {\em mass loss dispersion} ($\sigma_M$), its interplay with the 
effects of second parameters acting on the {\em mean ZAHB mass} 
($\langle M_{\rm HB}\rangle$), 
i.e. age, metal content, etc., may significantly contribute to the 
puzzling variety of HB morphologies actually observed in the Galactic 
globular clusters system.
Since we still lack a 
detailed description of how dense stellar media might contribute to the 
occurrence of increased mass loss and thus to a bluer HB morphology, we 
are accordingly unable to take into due account 
the possible effects of the tremendous 
difference in central concentration between NGC~288 and NGC~362 in our 
age difference analysis. A deeper 
understanding of mass loss processes in red giant stars is badly needed 
in order for analyses like the one we have carried out in the present 
paper to be placed on a firmer footing, and for a {\em conclusive} answer  
to the second parameter problem to be provided.

\acknowledgments
We would like to thank Dana Dinescu, Frank Grundahl, Bob Rood and 
Don VandenBerg for insightful discussions and useful information. 
We gratefully acknowledge the useful suggestions by an anonymous referee. 
Support for M.C. was provided 
by NASA through Hubble Fellowship grant HF--01105.01--98A awarded by the 
Space Telescope Science Institute, which is operated by the Association
of Universities for Research in Astronomy, Inc., for NASA under
contract NAS~5--26555. M.B., F.F.P. and F.R.F. have been supported by the 
Italian MURST through the COFIN p. {\small 9902198923\_004} grant, 
assigned to the project 
{\em Stellar Dynamics and Stellar Evolution in Globular Clusters.} 
The financial support to S. Galleti has been provided by the 
Osservatorio Astronomico di Bologna.
M.C. is grateful to the staff of the Osservatorio Astronomico di Bologna, 
where a substantial part of this work was carried out, for its hospitality 
and generous support.

\end{document}